\documentstyle[11pt,newpasp,twoside,epsf]{article}
\def\edcomment#1{\iffalse\marginpar{\raggedright\sl#1\/}\else\relax\fi}
\reversemarginpar

\begin{document}
\title{Imaging of Radio Supernovae}
\author{Michael Bietenholz}
\affil{Dept.\ of Physics and Astronomy, York University, 4700 Keele St.,
Toronto, ON, M3J 1P3, Canada.}

\begin{abstract} 
As the shock and expanding shell of a supernova plow out through the
circumstellar material at thousands of km s$^{-1}$, radio emission is
generated.  VLBI observations of this radio emission are presently the
only means to directly image the expanding shell of any supernova much
farther away than SN~1987A.  The last decade has seen great progress
in VLBI imaging of radio supernovae.  In particular, SN~1993J in the
galaxy M81 provided a rare opportunity to closely study an expanding
supernova, and has been intensively observed.  I summarize some of the
results on SN~1993J and other radio supernovae, and compare the
different observed radio supernovae.  I briefly discuss the future
prospects of radio supernovae imaging.
\end{abstract}

\section{Introduction}

The first known supernova observed with VLBI was the type II supernova
1979C\index{SN1979C}, which was only the second
\index{supernova}\index{radio supernova}supernova detected in the
radio, and
Bartel (1985) reported a determination of its angular size. The first
VLBI image identified as that of a supernova was one of
\index{41.95+575} 41.95+575 in M82 by Wilkinson \& de Bruyn (1984),
although the recent results have cast some doubt on the identification
as a supernova (McDonald et al.\ 2001).  Bartel et al.\ (1987)
published a higher resolution image of 41.95+575, and then Bartel et
al.\ (1991) published an image of \index{SN1986J}SN~1986J\@.  Since
then, radio imaging of approximately ten supernovae has been carried
out (see Table~1 below), where I use the word ``imaging'' somewhat
loosely, meaning supernova for which we have some morphological
information, including size determinations, from radio observations.
I will attempt to briefly summarize the state of radio supernova (RSN)
imaging.

Radio imaging is an important tool for the study of supernovae, at
least those that are radio bright, ie., RSNe.  If fact,
\index{VLBI}VLBI radio observations are presently the only means to
directly image the expanding shell of any supernova much farther away
than \index{SN1987A}SN~1987A\@. The radio emission is generated as the
expanding shell of supernova ejecta interacts with the progenitor
star's circumstellar medium (CSM) which generally consists of its
stellar wind. The expanding shell of ejecta generates two shocks, a
forward shock which propagates into the CSM and a reverse shock which
propagates into the stellar ejecta.  Chevalier (1990 and references
therein) showed that in the case of density distributions for the CSM
and the ejecta that are power laws in radius, the solution is
self-similar and the supernova shell expands such that its radius, $r
\propto t^m$, where $t$ is time since the explosion and the parameter
$m$ is commonly known as the deceleration parameter.  The value of $m$
will depend on the density distributions of both the CSM and of the
ejecta, and $m = 1.0$ indicates free expansion.
\begin{figure}
\centering \leavevmode
\epsfxsize=0.60\textwidth \epsfbox{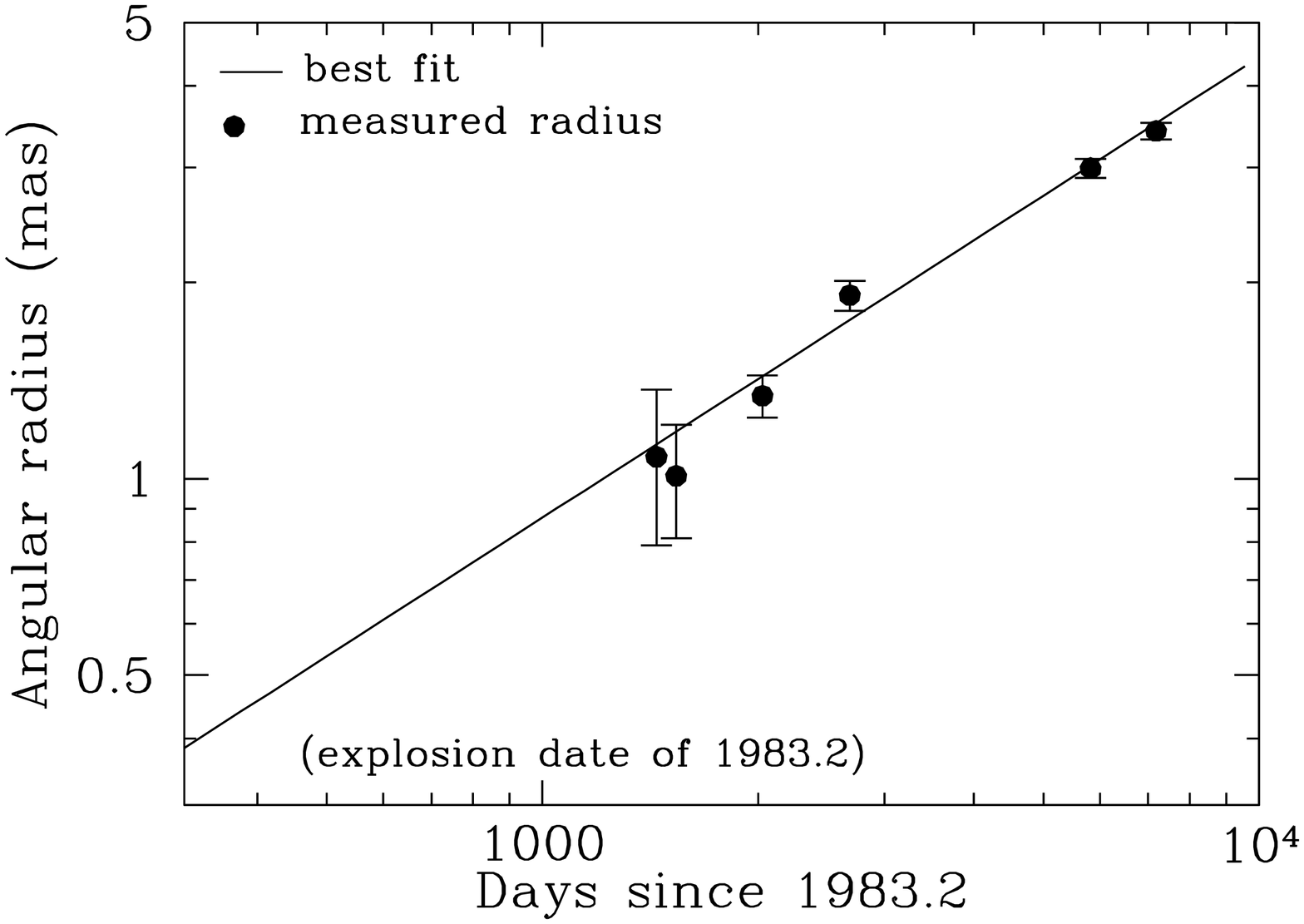}
\caption{The expansion curve of SN~1986J. The radii are average outer
radii measured in the image plane (see Bietenholz, Bartel, \& Rupen
2002).  The line indicates the best fit to all but the last
point, with angular radius $\propto t^{0.71}$. \label{bieten_f86exp}}
\end{figure}

\begin{figure}[th]
\centering \leavevmode
\epsfxsize=0.86\textwidth \epsfbox{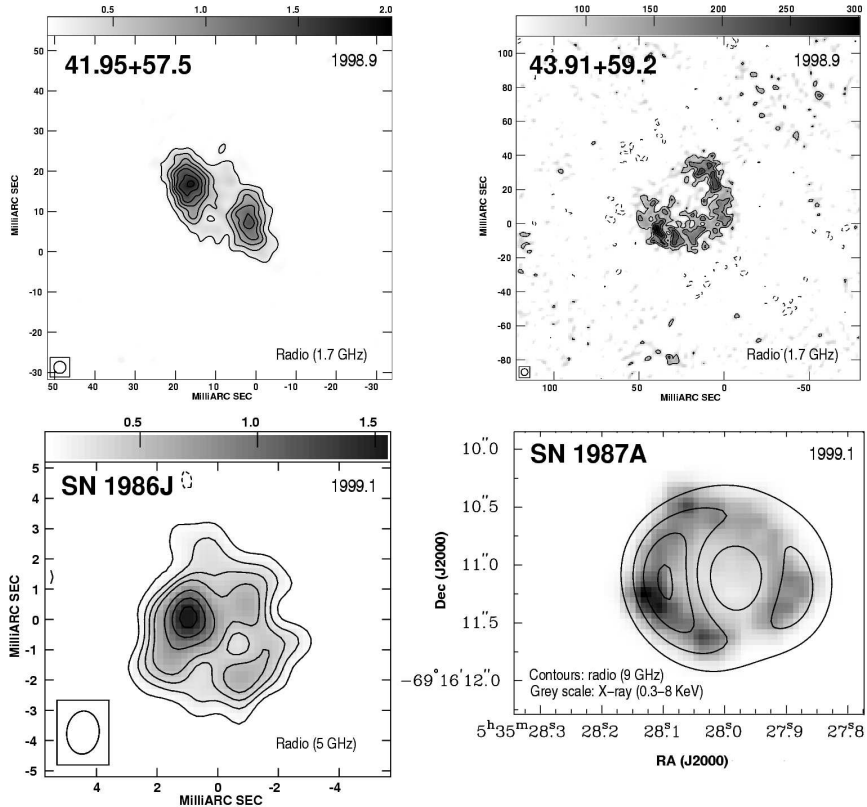}
\caption{Radio images of supernovae at the epochs given in the upper
right. {\bf 41.95+57.5}: (the identification as a supernova is
uncertain, McDonald et al.\ 2001). {\bf 43.91+59.2}: (McDonald et al.\
2001). {\bf SN~1986J}: (Bietenholz et al.\ 2002).  In the first three
panels, the contours and the grey scale, labeled in mJy~beam$^{-1}$, both
show the radio emission. {\bf SN~1987a}: The contours, at 1.1, 2.2,
3.3, and 4.4 mJy~beam$^{-1}$, show the radio emission and the grey scale
shows the X-ray emission (Manchester et al.\ 2002).
\label{bieten_fgallery}}
\end{figure}
The interaction region between the ejecta and the CSM is an important
laboratory for studying shock acceleration, since supernovae are
thought to be responsible for the majority of cosmic rays.  We can
also learn about both the supernova and the CSM\@.  Furthermore, since
the speeds of the ejecta are usually several orders of magnitude faster
than those typical of red giant winds, the supernova shock over-runs
the wind, and thus the interaction region acts like a time-machine,
and has the potential to reveal the last thousands of years of the
progenitor's wind history.  Finally, by equating the angular expansion
velocity measured with VLBI with the radial expansion velocity
measured spectroscopically, a direct, geometric distance determination
can be made (see Bartel \& Bietenholz, this volume; Bartel 1985).

The early VLBI observations of SN~1979C and SN~1986J allowed
measurement of their sizes and determination of their expansion
curves.  Figure~\ref{bieten_f86exp} shows the expansion curve to date
of SN~1986J (see Bietenholz, Bartel, \& Rupen 2002; for SN~1979C, see
Bartel \& Bietenholz, this volume). A power law, as suggested by
the Chevalier mini-shell model, indeed provides a good fit to the
data, with $r \propto t^{\,0.71}$ (but see P\'erez-Torres et al.\ 2002
for a different result).  A radio image of SN~1986J is shown in
Figure~\ref{bieten_fgallery}.

Several decades-old supernovae have been observed in \index{M82}M82
(see e.g., McDonald et al.\ 2001), and images of two are shown in
Figure~\ref{bieten_fgallery}.  For \index{43.31+592}43.31+592, $m$
could be determined to be 0.7.  Radio images of SN~1987A have been
obtained using the Australia Telescope (see Manchester et al.\ 2002)
and are also shown in Figure~\ref{bieten_fgallery}.  Its expansion can
be shown to have been strongly decelerated.

\begin{figure}[th]
\centering \leavevmode
\epsfxsize=0.85\textwidth \epsfbox{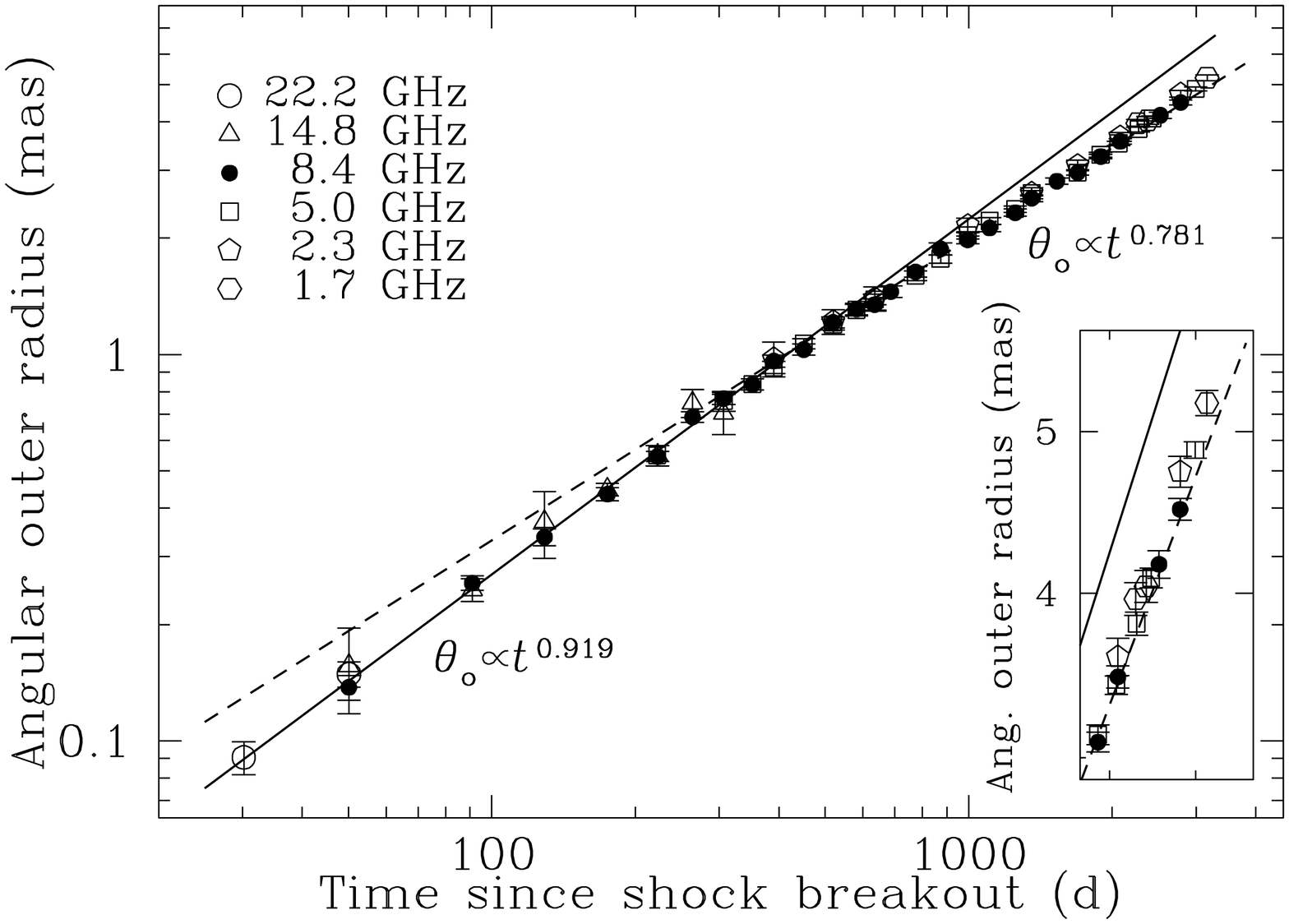}
\caption{The expansion curve of SN~1993J (Bartel et al.\ 2002). The
angular outer radii, $\theta_{\rm o}$, are derived from a spherical
shell model fit directly to the visibility data at frequencies between
1.7 and 22.2~GHz.  The solid and dashed lines are power-law fits to the
expansion, showing a change in the power-law index, $m$, from 0.919 to
0.781 at $t \sim 600$~d.  The inset shows the outer radii for the
latest epochs in more detail, showing the subsequent increase of $m$
after $t \sim 1600$~d.
\label{bieten_f93exp}}
\end{figure}
\begin{figure}[p]
\centering 
\epsfxsize=0.93\textwidth \epsfbox{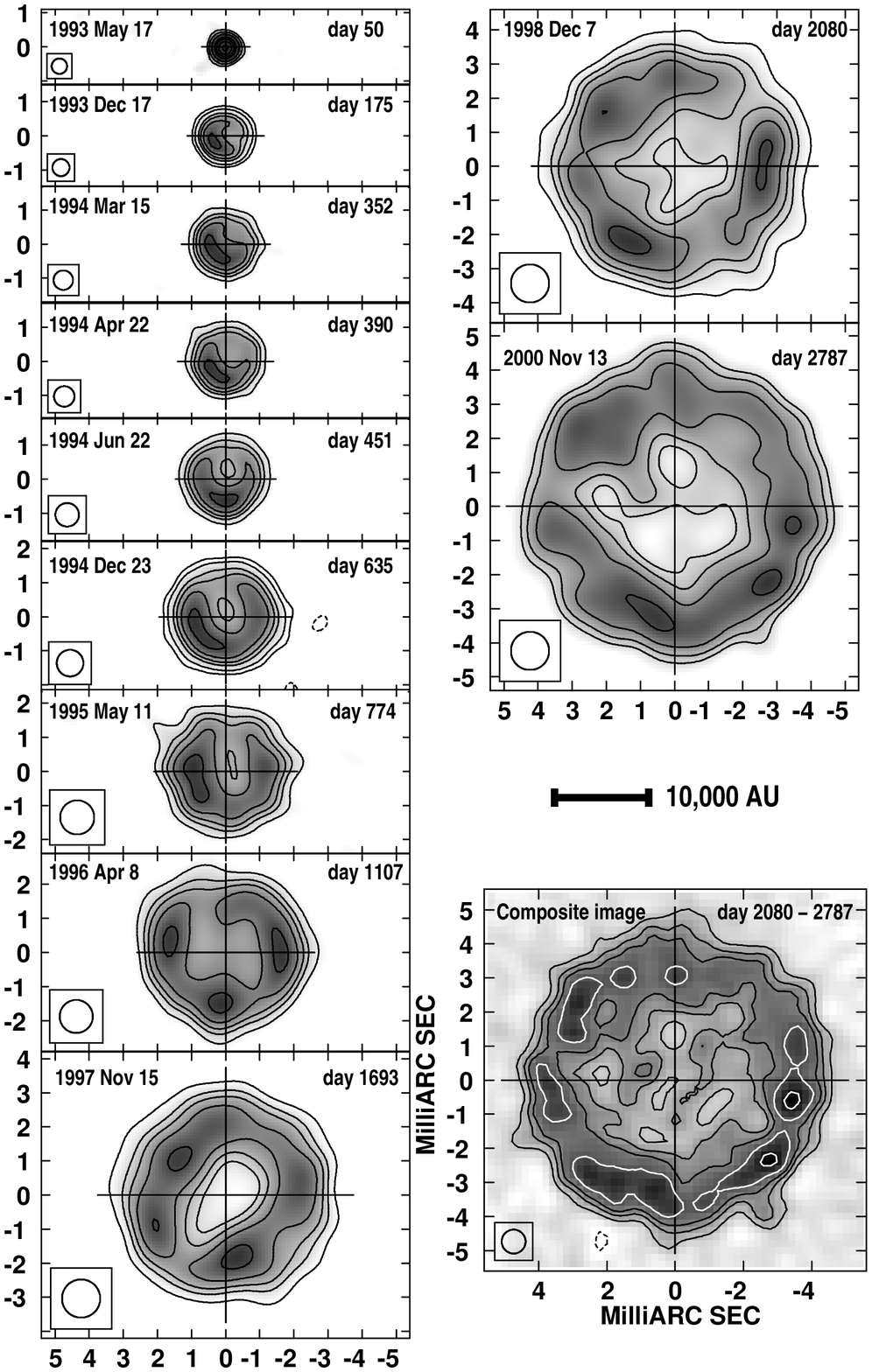}
\caption{\small VLBI images of SN~1993J at 8.4~GHz.  The axes are
labeled in mas and the FWHM resolution is shown at lower left in each
panel.  The last panel shows a higher-resolution composite image, made
by coherently averaging the data from 2080 to 2787~d after the
explosion.   The contours are drawn at 1, 2, 4, 8, 16, 32, 45, 64, and
90\% of the peak brightness, but only if they are $> 3\times$ the rms
background noise (see Bietenholz et al.\ 2003).
\label{bieten_f93img}}
\end{figure}
By far the best studied RSN so far is \index{SN1993J}SN~1993J in the
nearby spiral galaxy \index{M81}M81.  Two independent VLBI campaigns
have observed it extensively (see Bie\-tenholz, Bartel, \& Rupen 2003;
Bartel et al.\ 2000; and Marcaide et al.\ 2002b, 1997; and references
therein).  The quality of the data allowed both VLBI groups to
independently determine departures from simple power-law expansion,
with Bartel et al.\ (2002) reporting that $m(t)$ was near unity early
on, decreased to $\la\,$0.8 by $t=1600$~d, and then increased slightly
after that (see Figure~\ref{bieten_f93exp}).

Mioduszewski, Dwarkadas, \& Ball (2001) performed simulations of an
expanding supernova shell, based on a particular model of the ejecta
profile, and found changes in $m(t)$ remarkably similar to those in
fact observed for SN~1993J (e.g., Bartel et al.\ 2002).  After 10
years, SN~1993J has slowed to less than half its original expansion
velocity, which implies that the swept up material is now comparable
in mass to the ejecta, and for reasonable mass-loss rates of the
progenitor, suggests that the ejecta have a relatively low mass of
$\sim 0.3 \, M_{\sun}$, and that the progenitor had lost most its
envelope mass to a binary companion prior to the supernova explosion.

The images of SN~1993J, shown in Figure~\ref{bieten_f93img} show a
wealth of detail (see {\tt
http://www.yorku.ca/bar\-tel/\-SNmovie.html} for a movie of the
expanding supernova).
In projection, the supernova has remained very circular, and phase
referencing has shown that it expands isotropically from the explosion
center to within 5.5\% (Bietenholz, et al.\ 2001, 2003).  The shell
structure is clearly visible in all the adequately resolved images.
There are, however, distinct departures from circular symmetry even at
the earliest resolved epochs, where there is a pronounced minimum to
the west and a maximum to the east-southeast
(Fig.~\ref{bieten_f93img}).  Even after 10 years, no sign of a
\index{pulsar nebula}pulsar nebula has yet been seen at the center.
However, it seems probable that the ejecta in the interior of the
shell are dense enough that the radio optical depth due to free-free
absorption will remain large for several decades, and hence obscure
any emerging pulsar nebula.

\section{Conclusions}

\begin{table}[ht]
\begin{minipage}{\textwidth}
\footnotesize
\begin{tabular}{l l c c c c l c}
\tableline
 Name &	Host   &Distance&Type& Expansion & $m^{\rm a}$
    & Morphology & Ref.$^{\rm b}$ \\
      & galaxy &          &  & speed  &  \\
      &        & (Mpc)    &   &($10^3$ km~s$^{-1}$) &  \\
\tableline  
SN 1979C	  & M100    & 16\phantom{0} 
                          & II   &\phantom{55}12 & 0.95 &	circular & 3 \\
SN 1980K	  & NGC6946 & 6   & II   & $<22$         &      &            & 1 \\
SN 1986J   & NGC891  & 10\phantom{0} 
                          & II   &\phantom{555}6 & 0.7\phantom{1}& distorted shell & 5 \\
41.95+575$^{\rm c}$& M82	    & 3   &      &\phantom{5}$<2$&      & asymmetric  & 7 \\
43.31+592 & M82	    & 3   &      &\phantom{55555}9.7   
                                                 & 0.7\phantom{1}& circular shell & 7 \\
44.01+596$^{\rm c}$& M82$^{\rm d}$& 3   & 	 &               &      & circular shell & 7 \\
SN 1987A	  & LMC     & \phantom{.05}0.05
                          &IIpec &\phantom{5}$\sim 3$ & 
       $<0.3$\phantom{111}& circular shell & 6 \\
SN 1993J	  & M81     & 4   & II   &\phantom{555.5}8.5 
                                                 & 0.81 & circular shell & 2 \\
SN 1994I	  & M51	    & 8   & Ic   & $<65$         &      &    &  4 \\
SN 2001gd  & NGC5033 & 13\phantom{0}  & II & $\sim 25$ 
                                                 &      &    &  8 \\
\tableline
\end{tabular}
\tablenotetext{a}{Average value of the deceleration parameter, $m$.}
\tablenotetext{b}{References (e.g.): 
 1 = Bartel 1985; 
 2 = Bartel et al.\ 2002, Marcaide et al.\ 2002b;
 3 = Bartel \& Bietenholz 2003; 
 4 = Bartel \& Bietenholz, unpublished data; 
 5 = Bietenholz, Bartel, \& Rupen 2002; 
 6 = Manchester et al.\ 2002;
 7 = McDonald et al.\ 2001; 
 8 = P\'erez-Torres et al.\ 2003.}
\tablenotetext{c}{Identification as a supernova/supernova remnant uncertain.}
\tablenotetext{d}{We list only the most prominent sources in M82.
Some information is available on others, see Bartel et al.\ 1987,
Pedlar et al.\ 1999.}
\end{minipage}
\end{table}

Table~1 gives an overview of the characteristics of
the RSNe for which we have some morphological information.  Of the
well-resolved supernovae, all but 41.95+575, whose identification as a
supernova is uncertain, show a discernible, fairly circular shell
morphology.  The observed shell structures are clear evidence that the
bulk of the radio emission originates from the interaction of the
expanding shells with their surroundings, rather than from nebulae
around the young pulsars, as had also been suggested.  Despite the
relatively circular shells, significant intensity modulation with
position angle seems the norm.  None of the imaged RSNe show a clear
case of bisymmetric structure, as might be expected from axisymmetric
stellar winds.  There is evidence, however, most compelling from
SN~1986J, for some more asymmetric structure, possibly in the form of
protrusions or jets.  The origin of the asymmetries and protrusions is
not yet well understood.  The observed mostly circular morphologies,
do, however, suggest mostly spherical shells, which allow direct
distance determinations to be made (e.g., Bartel \& Bietenholz, this
volume; Bartel 1985).

Power-law expansion curves, with $r \propto t^{\,0.7 \sim 1.0}$, provide
a first order fit to all supernovae\footnote{Although see Marcaide,
et~al.\ 2002a, for an alternate view of SN~1979C} whose sizes have
been determined over a reasonable time baseline except for SN~1987A,
which is probably in a much lower density environment than the other
radio-bright supernovae.  However, at least in the case of SN~1993J,
distinct departures from strict power-law expansion can be seen,
implying significant structure in the ejecta and/or the CSM\@.  In
both \index{SN1993J}SN~1993J and \index{SN1986J}SN~1986J, the source
structure has been shown to change non self-similarly as the supernova
expands.  There is an intriguing change in the radio spectrum of
SN~1986J, with a new spectral component that peaks at $\sim$20~GHz
emerging after $\sim$10~yr, possibly related to a new compact
component in the images, that may perhaps be related to a pulsar
nebula (see Bietenholz et al.\ 2002).

There is still considerable scope for the more detailed understanding
of supernova shell dynamics, and for determining the structure of
the CSM and of the ejecta through hydrodynamic simulations,
particularly in the case of SN~1993J where exceptionally good data are
on hand.

\section{The Future of Imaging of Radio Supernovae}

Currently, there are only a couple of \index{radio supernova}RSNe per
decade that are close enough and bright enough to image.  Future
improvements in sensitivity, e.g., the Mark~IV VLBI recording system,
the eVLA, and the SKA, will allow us to image somewhat more RSNe, but
will only enable a modest increase in the resolution available to
global VLBI (by going to higher observing frequencies).  The practical
distance limit is on the order of 30~Mpc, where global VLBI at 22~GHz
gives a linear resolution of $\sim\,$8000~AU\@.  There have been
approximately 8 RSNe per decade with a 22-GHz flux density $> 0.1$~mJy
since 1980, so even with the SKA, the number of RSNe available for
imaging will be limited.

It will therefore be important to image those RSNe that are available
to us, since they will provide us with direct distance estimates, test
our understanding of the supernova process, which may be crucial to
understanding the distant, un-resolved supernovae upon which
cosmological results are based, and allow us to probe the
circumstellar environments and thus the wind-history of evolved stars,
and allow us maybe to see the birth of a pulsar nebula.

\end{document}